\documentclass[letterpaper,journal]{IEEEtran}
\usepackage{amsmath,amsfonts}
\usepackage{algorithmic}
\usepackage{algorithm}
\usepackage{array}
\usepackage[caption=false,font=normalsize,labelfont=sf,textfont=sf]{subfig}
\usepackage{threeparttable}
\usepackage{textcomp}
\usepackage{stfloats}
\usepackage{url}
\usepackage{verbatim}
\usepackage{graphicx}
\usepackage{cite}
\usepackage{multicol}
\usepackage{multirow}
\usepackage{booktabs}
\usepackage{makecell}
\usepackage{pifont}
\usepackage[table]{xcolor} 
\usepackage{color}
\definecolor{rowcolor}{rgb}{0.898, 0.949, 0.969}
\usepackage{tcolorbox}
\usepackage[colorlinks,
linkcolor=blue,
anchorcolor=black,
citecolor=black]{hyperref}
\usepackage{bbm}
\definecolor{color4audio}{rgb}{0.71, 0.96, 0.91}
\definecolor{color4emg}{rgb}{0.97, 0.76, 0.34}
\definecolor{color4vis}{rgb}{0.53, 0.8, 0.31}
\definecolor{color4fus}{rgb}{0.39, 0.39, 0.39}

\begin{document}
	\title{AVE Speech: A Comprehensive Multi-Modal Dataset for Speech Recognition Integrating Audio, Visual, and Electromyographic Signals}
	
	\author{Dongliang~Zhou, Yakun~Zhang,
		Jinghan~Wu, Xingyu~Zhang, Liang~Xie, and Erwei~Yin
		\thanks{This work was supported in part by the National Natural Science Foundation of China under Grant 62332019, the National Key Research and Development Program of China (Grants 2023YFF1203900 and 2023YFF1203903), the Beijing Nova Program (Grant 20240484513),  and the open project of Sichuan Provincial Key Laboratory of Philosophy and Social Science for Language Intelligence in Special Education (No. YYZN-2025-1).
			D. Zhou is with Defense Innovation Institute, Academy of Military Sciences, Beijing, China;  Tianjin Artificial Intelligence Innovation Center, Tianjin, China; and Harbin Institute of Technology, Shenzhen, China. Y. Zhang, X. Zhang, L. Xie, and E. Yin are with Defense Innovation Institute, Academy of Military Sciences, Beijing, China, and Tianjin Artificial Intelligence Innovation Center, Tianjin, China. J. Wu is with Tianjin University, Tianjin, China. Corresponding author: Erwei Yin, e-mail: yinerwei1985@gmail.com.}
	}

	\maketitle
	\begin{abstract}
		The global aging population faces considerable challenges, particularly in communication, due to the prevalence of hearing and speech impairments. To address these, we introduce the AVE speech, a comprehensive multi-modal dataset for speech recognition tasks. The dataset includes a 100-sentence Mandarin corpus with audio signals, lip-region video recordings, and six-channel electromyography (EMG) data, collected from 100 participants. Each subject read the entire corpus ten times, with each sentence averaging approximately two seconds in duration, resulting in over 55 hours of multi-modal speech data per modality. Experiments demonstrate that combining these modalities significantly improves recognition performance, particularly in cross-subject and high-noise environments. To our knowledge, this is the first publicly available sentence-level dataset integrating these three modalities for large-scale Mandarin speech recognition. We expect this dataset to drive advancements in both acoustic and non-acoustic speech recognition research, enhancing cross-modal learning and human-machine interaction.
	\end{abstract}
	
	
	\section{Introduction}
	
	\IEEEPARstart{T}{he} challenges associated with an aging global population have increasingly become a focal point of societal concern. According to a report\footnote{\url{https://www.who.int/news-room/fact-sheets/detail/ageing-and-health}} by the World Health Organization (WHO), the proportion of individuals aged 60 and above is projected to nearly double, from 12\% in 2015 to 22\% by 2050. This demographic shift imposes significant demands on health and social care systems worldwide, necessitating their adaptation to meet the evolving needs of the elderly population. One of the key challenges faced by the elderly is the difficulty in articulating their needs clearly due to hearing or speech impairments, which can hinder effective communication. Thus, the timely and accurate recognition of these needs is of paramount importance. In this context, speech recognition technologies emerge as critical tools, offering a natural and efficient means of facilitating human-machine interaction \cite{9508143}.
	\begin{figure}[t]
		\centering
		\includegraphics[width=0.5\textwidth]{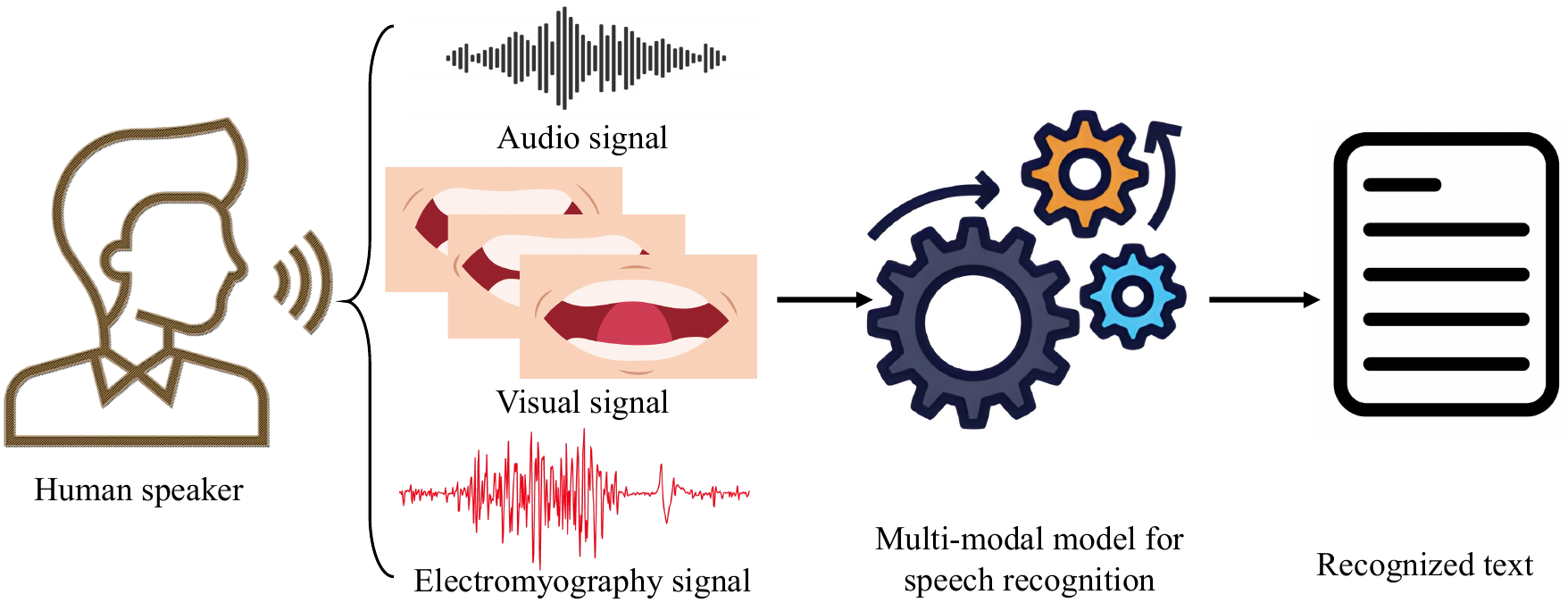}
		\caption{Illustration of multi-modal speech recognition pipeline integrating audio, visual, and electromyographic signals.}
		\label{fig:cover}
	\end{figure}
	Speech recognition can generally be categorized into two main approaches: automatic speech recognition (ASR) \cite{trentin2003robust,bahdanau2016end,schneider2019wav2vec,baevski2020wav2vec,hsu2021hubert,wang2023complex,radford2023robust} using audio signals and silent speech recognition (SSR) \cite{matthews2002extraction,zhang2023emg,sun2024earssr,hao2025lipgen} using non-acoustic signals. These advancements are poised to revolutionize consumer electronics, making them indispensable in enhancing the quality of life for the aging population. ASR has become deeply integrated into daily life, exemplified by the speech-to-text functionality in numerous mobile applications such as Apple's Siri and Xiaomi's Xiao Ai. Despite the widespread utility and impressive capabilities of ASR, its accuracy significantly diminishes in environments with high noise levels or in situations where acoustic signals cannot be reliably captured. To address these limitations, SSR has seen accelerated development, particularly in sectors like elderly care, disability assistance, and environments with significant background noise. The progress in SSR has been further propelled by advancements in training datasets and deep learning algorithms, leading to substantial improvements in specific areas such as visual speech recognition (commonly known as lipreading) \cite{matthews2002extraction} and surface electromyography (EMG)-based speech recognition \cite{wand2014tackling}. However, visual speech recognition faces challenges such as sensitivity to lighting conditions and the inherent difficulty in distinguishing between words with similar mouth shapes. Similarly, surface EMG is not without its own limitations: it is susceptible to variability due to physiological differences among users, and the positioning of electrodes can shift during speech, potentially compromising signal integrity. To overcome these individual limitations, integrating multiple modalities can offer complementary information that enhances the robustness and accuracy of speech recognition, thereby enabling unconstrained and more reliable communication. As shown in Fig. \ref{fig:cover}, leveraging the strengths of each modality allows this multi-modal system to effectively overcome the limitations inherent in single-modal speech recognition systems, providing more robust and accurate communication solutions.

    \begin{table*}[t]
		\centering
		\caption{Comparison of Speech Recognition Datasets. Here, `\#' indicates the count of the specified metric}
		\label{dataset}
		\resizebox{\linewidth}{!}{
			\begin{tabular}{lccccccc}
				\toprule
				\textbf{Dataset}   & \textbf{Modality}          & \textbf{Language}     & \textbf{Corpus content}        & 
                \textbf{Data collection context}
                & \textbf{\# Speakers} & \textbf{Speaker ID} & \textbf{Interaction focus}\\
				\midrule
				LibriSpeech \cite{panayotov2015librispeech}    & Audio            & English     & 1,000 hours of spoken sentences                                        & 
                Online collection of book reading
                & 1,000 & \ding{55}   &  \ding{55}     \\
				THCHS30 \cite{wang2015thchs}      & Audio            & 
                Mandarin
                & 1,000 sentences                                                   & On-site collection of news sentences   & 40   & \ding{55}  & \ding{55}  \\
				LRW \cite{chung2017lip}       & Visual           & English     & 500 classes of words                                                & Online collection of news broadcasts         & 1,000 & \ding{55}  &  \ding{55}      \\ 
				LRW-1000 \cite{yang2019lrw}     & Visual           & 
                Mandarin
                & 1,000 classes of words  & Online collection of news broadcasts       & 2,000 & \ding{55}    & \ding{55}   \\
				LRS2-BBC \cite{afouras2018deep}     & Audio and visual        & English     & Thousands of sentences                          &  Online collection of news broadcasts
                & --     & \ding{55}   &  \ding{55}     \\
				EMG-UKA \cite{wand2014emg}     & Audio and EMG         & English     & More than 7,000 utterances                                             &  On-site collection of news sentences         & 8         & \ding{51} & \ding{55}        \\
				\midrule
				\rowcolor{rowcolor}
				AVE speech (\textbf{ours}) & \textbf{Audio, visual, and EMG} & 
                Mandarin
                & 100 classes of sentences  & \raisebox{-0.2\height}{\shortstack{On-site collection of assistive \\daily and medical sentences}}    & 100    & \textbf{\ding{51}}  &  \textbf{\ding{51}}  \\ 
				\bottomrule
		\end{tabular}}
	\end{table*}
    
	The development of advanced speech recognition systems relies heavily on high-quality datasets that encompass diverse speech modalities. Open-source datasets have played a crucial role in advancing speech recognition research, as exemplified by several examples in Table \ref{dataset}. Widely accessible audio datasets, such as LibriSpeech \cite{panayotov2015librispeech} and THCHS30 \cite{wang2015thchs}, have been instrumental in this regard. 
    Multi-modal datasets combining audio and visual modalities, such as the LRS2-BBC dataset by Afouras \textit{et al.} \cite{afouras2018deep}, have further advanced multi-modal speech recognition algorithms. Additionally, EMG-based datasets, exemplified by EMG-UKA \cite{wand2014emg}, have explored physiological signals coupled with audio, although these datasets typically involve limited speaker diversity.
   Despite these advancements, no existing dataset comprehensively integrates audio, visual, and EMG modalities with sufficient speaker diversity and detailed speaker information, particularly addressing the practical challenges faced by elderly individuals and those with communication impairments.
	
	Building upon the need to address challenges faced by the elderly and individuals with communication impairments, integrating audio, visual, and EMG signals into a multi-modal speech recognition system holds significant potential. This approach can be applied to a broader range of scenarios, including rehabilitation for patients with speech disorders, daily assistance for the elderly, and private communication in low-light or dynamic indoor environments. To address the limitations associated with single-modality speech recognition—particularly the lack of large-scale, comprehensive datasets that include multiple modalities and detailed speaker information—this paper introduces the AVE speech dataset. Designed specifically for multi-modal speech recognition, the AVE speech dataset integrates audio, visual, and EMG signals directly related to the speech process and provides a fusion paradigm for this emerging field. The AVE speech dataset will be accessible at \url{https://huggingface.co/datasets/MML-Group/AVE-Speech}.
	
	The main contributions of this research can be summarized as follows:
	\begin{enumerate}
		\item To the best of our knowledge, this is the first public dataset that synchronously acquires audio signals, lip image sequences, and facial EMG signals.
		\item Our dataset is the first sentence-level Mandarin corpus encompassing 100 daily life sentences, each containing three to five distinct words.
		\item Our dataset is the first multi-modal speech dataset constructed in a controlled on-site environment with 100 Mandarin subjects, rather than being compiled and annotated from public videos.
	\end{enumerate}
	
	The remainder of this paper is organized as follows. Section \ref{sec:related_work} briefly reviews works related to speech recognition. Section \ref{sec:ave_dataset} outlines the construction process of our proposed AVE speech dataset. In Section \ref{sec:Experiments}, we present a comprehensive set of experiments conducted to validate the effectiveness of the dataset. Section \ref{sec:limi_and_disc} discusses the dataset's limitations and outlines potential directions for improvement. Finally, Section \ref{sec:cnc} concludes the paper and offers suggestions for future research directions.

	\section{Related Work}
	\label{sec:related_work}
	This research falls into the field of speech recognition, which has a large existing body of literature. In this section, we review related works on automatic speech recognition (ASR), silent speech recognition (SSR), and datasets in speech recognition. We also highlight the features of this research in comparison to those of prior works.
	
	\textbf{Automatic Speech Recognition.} ASR \cite{trentin2003robust,bahdanau2016end,schneider2019wav2vec,baevski2020wav2vec,hsu2021hubert,wang2023complex,radford2023robust} systems are designed to process audio signals or direct speech inputs from microphones, converting them into text, ideally in the script corresponding to the spoken language. The foundational work by Trentin \textit{et al.} \cite{trentin2003robust} proposed a neural network-based model as a novel approach to acoustic modeling. Further, Bahdanau \textit{et al.} \cite{bahdanau2016end} incorporated an attention mechanism \cite{9705489,zhou2024learning} with a recurrent neural network (RNN) to learn the alignments between sequences of input frames and output labels. 
	Schneider \textit{et al.} \cite{schneider2019wav2vec} proposed the unsupervised pretraining of convolutional neural networks to improve ASR performance, introducing a noise contrastive binary classification task that enabled wav2vec to leverage large-scale unlabeled data. Subsequently, wav2vec 2.0 \cite{baevski2020wav2vec} was developed to identify fundamental speech units for self-supervised tasks, optimizing both the prediction of speech units and the learning of task-specific speech modeling. Advancing self-supervised speech representation further, HuBERT \cite{hsu2021hubert} combines acoustic features with long-range contextual information via offline clustering, outperforming wav2vec 2.0 in continuous speech recognition tasks.
	More recently, Wang \textit{et al.} \cite{wang2023complex} utilized a spiking transformer \cite{vaswani2017attention} to improve the performance of ASR. The multi-modal dataset proposed in this paper offers a valuable resource for ASR, specifically within the single acoustic modality.
	Around the same time, Radford \textit{et al.} \cite{radford2023robust} developed the Whisper speech processing system, which was pre-trained on extensive audio transcripts to improve speech recognition performance for specific datasets without requiring fine-tuning.
	
	\textbf{Silent Speech Recognition.} Unlike ASR, SSR \cite{matthews2002extraction,zhang2023emg,sun2024earssr} focuses on interpreting non-acoustic signals from a human speaker, serving as an alternative in scenarios where vocalization is either impractical, such as during a meeting, or ineffective, such as in a noisy environment. SSR primarily encompasses visual speech recognition \cite{10472054} and bio-signal-based speech recognition \cite{zhang2023emg}. Visual speech recognition, also called lipreading, plays a critical role in SSR, leveraging lip-region image sequences as visual cues to decode speech content. Matthews \textit{et al.} \cite{matthews2002extraction} pioneered the use of hidden Markov models (HMMs) to process visual speech information for speech recognition. Recently, Chang \textit{et al.} \cite{chang2024conformer} utilized a linear visual front-end in conjunction with a convolution-involved transformer architecture, conformer, to develop a simple yet efficient visual speech recognition framework, achieving state-of-the-art performance for continuous speech recognition. Bio-signals have also been successfully implemented in SSR, with electromyography (EMG)-based speech recognition demonstrating notable performance and application potential. Initially, Lee \textit{et al.} \cite{lee2008emg} applied an HMM framework to the EMG-based recognition of 60 isolated words using three articulatory facial muscles. Subsequently, Wand \textit{et al.} \cite{wand2016deep} replaced the Gaussian mixture model (GMM) frontend with a deep neural network for continuous speech recognition using six-channel EMG signals. In a recent study, Zhang \textit{et al.} \cite{zhang2023emg} explored the cross-subject speech recognition approach with EMG signals, employing a convolutional neural network (CNN) \cite{zhou2023coutfitgan,10529644} as the recognition model and significantly extending the real-world application potential of EMG-based SSR. The AVE speech dataset proposed in this work includes both visual and EMG modality signals, designed to meet the specific requirements of SSR.
	
	
	\textbf{Datasets in Speech Recognition.}  Over the years, numerous datasets have been meticulously curated, each making a unique contribution to ASR and SSR research. Open-source datasets have significantly driven impact in speech recognition. Widely accessible audio datasets, such as LibriSpeech \cite{panayotov2015librispeech}, have been pivotal in this regard. SSR datasets have also been instrumental in algorithm development. For instance, Chuang \textit{et al.} \cite{chung2017lip} provided a substantial lipreading dataset comprising a 500-word corpus gathered from television broadcasts. In a subsequent study, Yang \textit{et al.} \cite{yang2019lrw} presented a large-scale Mandarin
    dataset for lipreading. Multi-modal speech recognition datasets have also been proposed. Afouras \textit{et al.} \cite{afouras2018deep} introduced and publicly released a large dataset for audio-visual speech recognition, LRS2-BBC. However, no previous work has covered three modalities of speech data with sufficient subject diversity when using physiological data. Moreover, existing large-scale multi-modal speech datasets, often constructed from television programs \cite{yang2019lrw}, lack speaker identity information, posing limitations for cross-subject research and the development of speaker-independent speech recognition systems. Our proposed AVE speech dataset, which includes audio, visual, and EMG modality signals, is designed to address these limitations, offering a comprehensive resource for the speech recognition research community.
	
	\section{AVE Speech Dataset}
	\label{sec:ave_dataset}
	The AVE speech dataset enables multi-modal speech recognition by integrating audio, visual, and EMG signals. This section details the design of the corpus, participant demographics, data collection equipment, dataset organization, data pre-processing, and basic feature extraction.
	
	\subsection{Corpus Design and Participants}
	\begin{table}[t]
		\centering
		\caption{Representative examples of sentences in the AVE speech corpus, categorized by the type of need. Each entry includes the English translation, the corresponding phonetic transcription in Mandarin, and the associated tonal patterns}
		\label{table:corpus}
		\resizebox{\linewidth}{!}{
			\begin{tabular}{lccc}
				\toprule
				\textbf{Type of need}     &  \textbf{\makecell{Sentence \\(English translation)}}           & \textbf{\makecell{Phonetic transcription \\(Mandarin)}} & \textbf{Tone sequence}   \\ \midrule
				Physiology       & I'm hungry               & wo e le              & 3 4 5   \\
				Safety         & Emergency               & jin ji hu jiu           & 3 2 1 4  \\
				Belongingness and love & I want to have a video chat      & wo yao liao shi pin        & 3 4 2 4 2 \\
				Esteem         & I can do it              & wo neng xing de          & 3 2 2 5  \\ \midrule
				Medical requirements  & I keep coughing            & wo yi zhi ke sou          & 3 4 2 2 4 \\
				Medical requirements  & My leg aches              & wo tui teng            & 3 3 2   \\
				Medical requirements  & How long do I need to stay in hospital & yao zhu yuan duo jiu        & 4 4 4 1 3 \\
				Medical requirements  & My blood pressure is high       & wo xue ya gao           & 3 3 1 1  \\ 
				\bottomrule
		\end{tabular}}
	\end{table}
	\begin{figure}[t]
		\centerline{\includegraphics[width=0.5\textwidth]{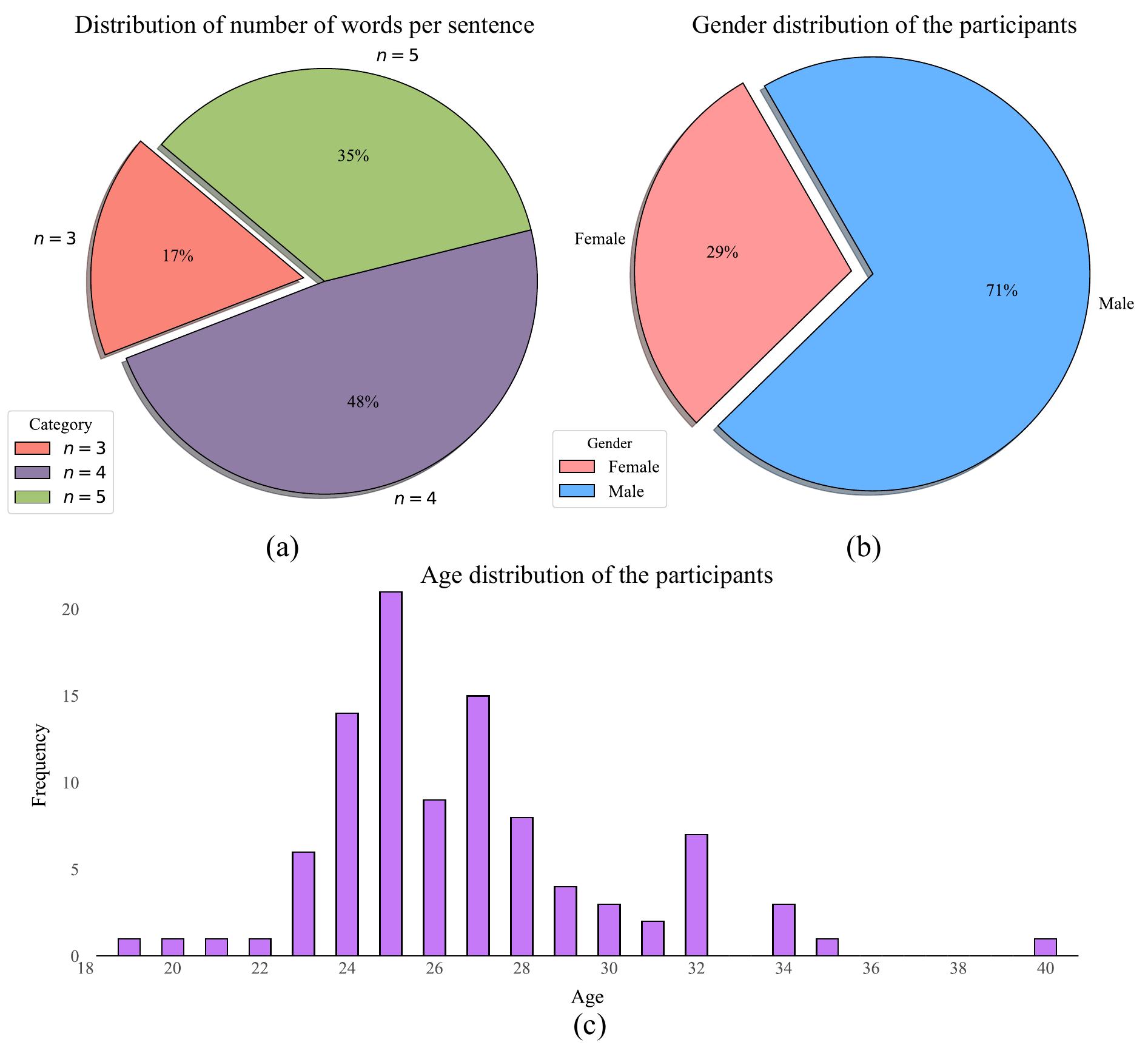}}
		\caption{Statistical overview of the corpus and participant demographics.}
		\label{dataset_stat}
	\end{figure}
	The AVE speech dataset is meticulously crafted based on the five levels of Maslow's Hierarchy of Needs: physiological, safety, belongingness and love, esteem, and self-actualization. Additionally, we incorporated sentences relevant to medical scenarios. In particular, the dataset comprises 100 Mandarin 
    sentences crafted to meet a broad spectrum of user needs, particularly those of individuals with speech impairments and elderly persons requiring assistance with daily living, rehabilitation, and caregiving. Each sentence in the corpus is an independent Mandarin expression, which is accompanied by phonetic transcriptions and tonal annotations, as outlined in Table \ref{table:corpus}. The tonal annotations correspond to the five distinct tones in Mandarin: the first tone (level tone), the second tone (rising tone), the third tone (dipping tone), the fourth tone (falling tone), and the fifth tone (neutral tone). For instance, the first sentence in Table \ref{table:corpus} represents a physiological need, translated into English as ``I'm hungry." The Mandarin pronunciation is ``wo e le," corresponding to the third, fourth, and fifth tones, respectively, for each 
    word. Another example is the fifth sentence, which pertains to medical requirements. This sentence, meaning ``I keep coughing," consists of five 
    words and is pronounced as ``wo yi zhi ke sou," with the tones arranged as the third, fourth, second, second, and fourth, respectively. It is worth noting that the data collection process was conducted entirely in Mandarin. English translations are provided solely for ease of understanding. The distribution of the number of words per sentence is illustrated in Fig. \ref{dataset_stat}(a), revealing that the sentences consist of three to five words. The entire corpus, including English translations and phonetic transcriptions, is available in the AVE speech dataset. Furthermore, the study involved 100 adult participants. The gender distribution, as shown in Fig. \ref{dataset_stat}(b), indicates that the participants consisted of 29 females and 71 males. As depicted in Fig. \ref{dataset_stat}(c), the participants ranged in age from 18 to 40 years, with a mean age of 26.68 years. All participants were native Mandarin speakers. The research was conducted with the approval of the Research Ethics Committee of our university, and informed consent was obtained from all participants prior to their involvement in the study.

\subsection{Data Collection Protocol}
	
\begin{figure*}[t]
		\centerline{\includegraphics[width=0.96\textwidth]{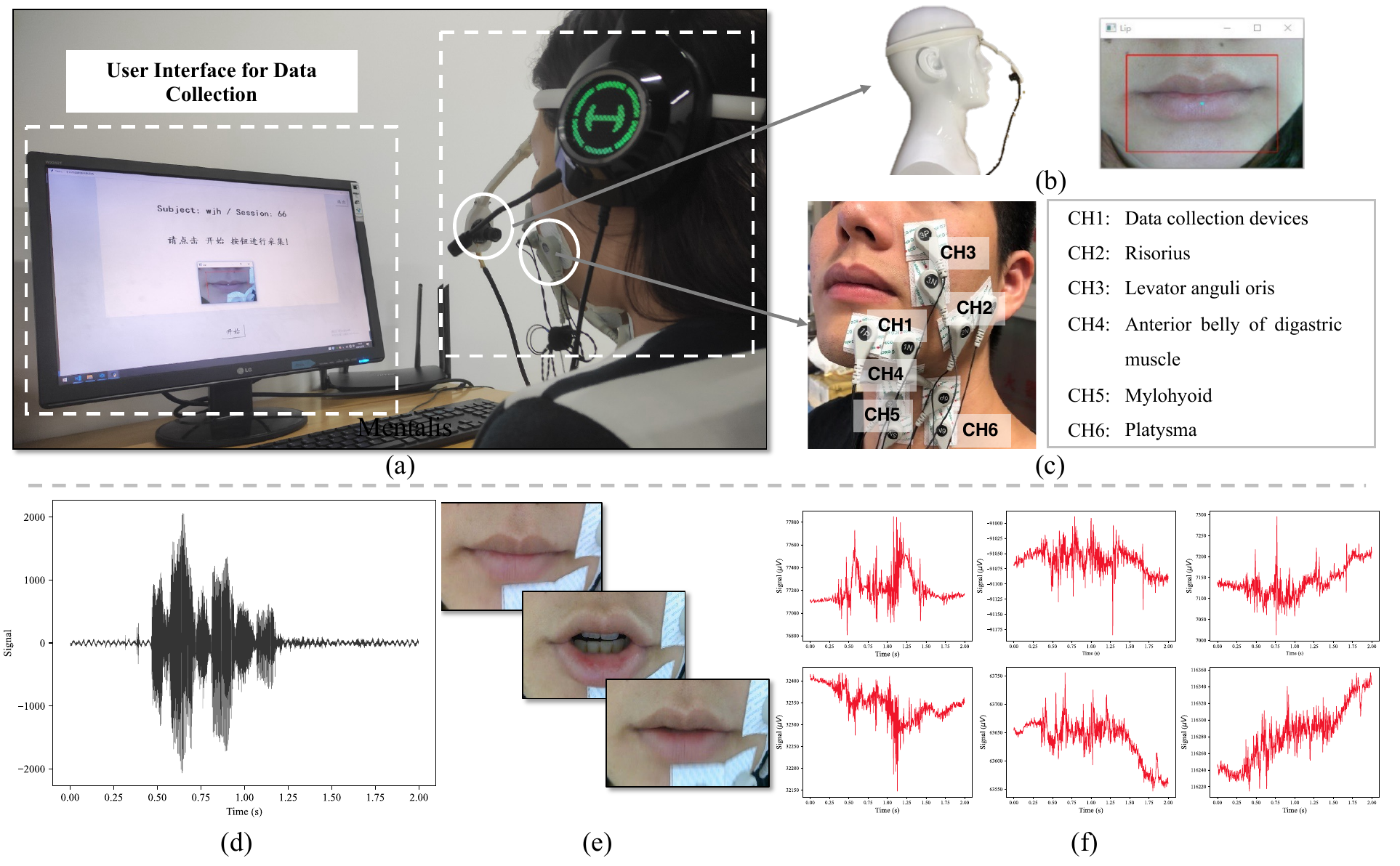}}
		\caption{Overview of the data collection system, including both hardware devices and the recording interface. (a) The user interface used for data collection, with the subject wearing the data collection devices. (b) The camera setup for capturing video of the lip region. (c) The locations of the six muscles from which surface EMG signals were recorded. (d) The waveform of the audio signal collected during speech. (e) Image sequence extracted from the lip-region video. (f) The waveform of the six-channel EMG signals recorded from the six muscles.}
		\label{fig:data-coll}
	\end{figure*}

This subsection details the data collection process, as depicted in Fig. \ref{fig:data-coll}. During the multi-modal data collection, participants were seated comfortably in a quiet room with normal lighting conditions. An interactive interface, illustrated in Fig. \ref{fig:data-coll}(a), provided detailed instructions to the participants. Once the participants donned the data collection devices, the collection process commenced. 
A prompt displayed on the interface instructed participants to read the presented sentences within two seconds. Participants were instructed to minimize unnecessary movements, such as head shaking, coughing, yawning, or swallowing, during the recording. Each complete reading of the corpus constituted one round of multi-modal data collection. 
Each participant completed ten rounds, with each round comprising 101 sentences. Within each round, a five-second break was provided after every 20 sentences. At the end of each round, participants were allowed to rest for several minutes until they felt ready to continue. The entire experiment lasted approximately one hour per participant. To mitigate potential biases from fixed sentence order or participant habituation, the sequence of the 101 sentences was randomly shuffled in each round for every participant.
For multi-modal speech data collection, a head-mounted microphone was used to record audio signals at a sampling rate of 44,100 Hz. An RGB camera was used to capture lip-region videos at a rate of 30 frames per second. The camera was positioned in front of the participant's lips, and the distance between the camera and the subject could be adjusted using a 3D-printed fixture, as shown in Fig. \ref{fig:data-coll}(b). A fixed boundary box was indicated in the center of the video image captured by the camera, with a size set to $640 \times 360$. The coordinates of the boundary points were then calculated. Surface EMG data were collected from six facial and neck muscles using the NSW308M bipolar EMG system (Neuracle Technology Co., Ltd), recording six-channel EMG data at a sampling rate of 1,000 Hz. Six pairs of electrodes were attached to the surface of these muscles, as depicted in Fig. \ref{fig:data-coll}(c). The first three channels recorded the EMG signals of the mentalis, risorius, and levator anguli oris muscles, which are located around the mouth region. These muscles influence the position of the lips during speech, resulting in variations in mouth shape corresponding to different speech contents. The fourth channel recorded the EMG signals from the anterior belly of the digastric muscle, located below the mandible, which lifts the hyoid bone during speech. The fifth and sixth channels captured the EMG signals from the mylohyoid and platysma muscles, located at the bottom of the mouth and neck, respectively. These muscles respond to the movement of the throat area during speech. A reference electrode was placed on the collarbone to record baseline voltage from the body. Electrode impedance was maintained below 10 $k\Omega\ $ during recording. In summary, audio signals, lip-region video images, and six-channel EMG signals corresponding to the same sentence were collected simultaneously, as shown in Figs. \ref{fig:data-coll}(d)-(f), respectively.
 
\subsection{Data Formats and Feature Processing}
\textbf{Data Formats.} The AVE speech dataset is meticulously organized to ensure effective data management and facilitate comprehensive research analysis. 
Detailed participant information, including subject index, gender, and age, is systematically compiled, ensuring that all pertinent data is readily accessible for analytical purposes. The dataset is structured hierarchically to facilitate ease of access and systematic analysis. At the highest level, the dataset is divided into three primary directories: audio, video, and EMG data. Each of these directories contains sub-directories corresponding to the 100 participants, forming the second level of organization. Within each participant’s directory, the data is further categorized by session, with each participant contributing data across ten sessions, constituting the third level of the structure. At the most granular level, each session directory houses 101 files, each representing single-modal speech data for an individual sentence, along with an additional file corresponding to a blank sentence. The files are systematically labeled, with comprehensive documentation of the labeling schema provided to ensure consistency in subsequent experimental use. During the data collection phase, 1,010 utterances were initially recorded from each participant. However, any data associated with errors during the reading of commands were excluded from the final dataset, resulting in 99,500 entries per modality, amounting to approximately 55.3 hours of recording. While basic data pre-processing can be applied to further enhance data quality, it is generally not recommended for the audio signals due to the high-quality collection achieved using a head-mounted microphone. For the video data, sequences of images are extracted from the lip region, converted to gray-scale, and then cropped to remove extraneous facial features. These images are subsequently resized to a standardized dimension of $88 \times 88$  pixels to ensure uniformity. The EMG data, however, may contain artifacts such as peak amplitude variations, DC offset, and 50 Hz power frequency noise, necessitating additional processing. To address these issues, a second-order Butterworth notch filter is employed to remove the 50 Hz noise and its odd harmonics at 250 Hz and 350 Hz. Additionally, a Butterworth band-pass filter, with a frequency range of 10 Hz to 400 Hz, is applied to retain the most relevant components of the EMG signals.
	\begin{figure*}[t]
		\centerline{\includegraphics[width=1\textwidth]{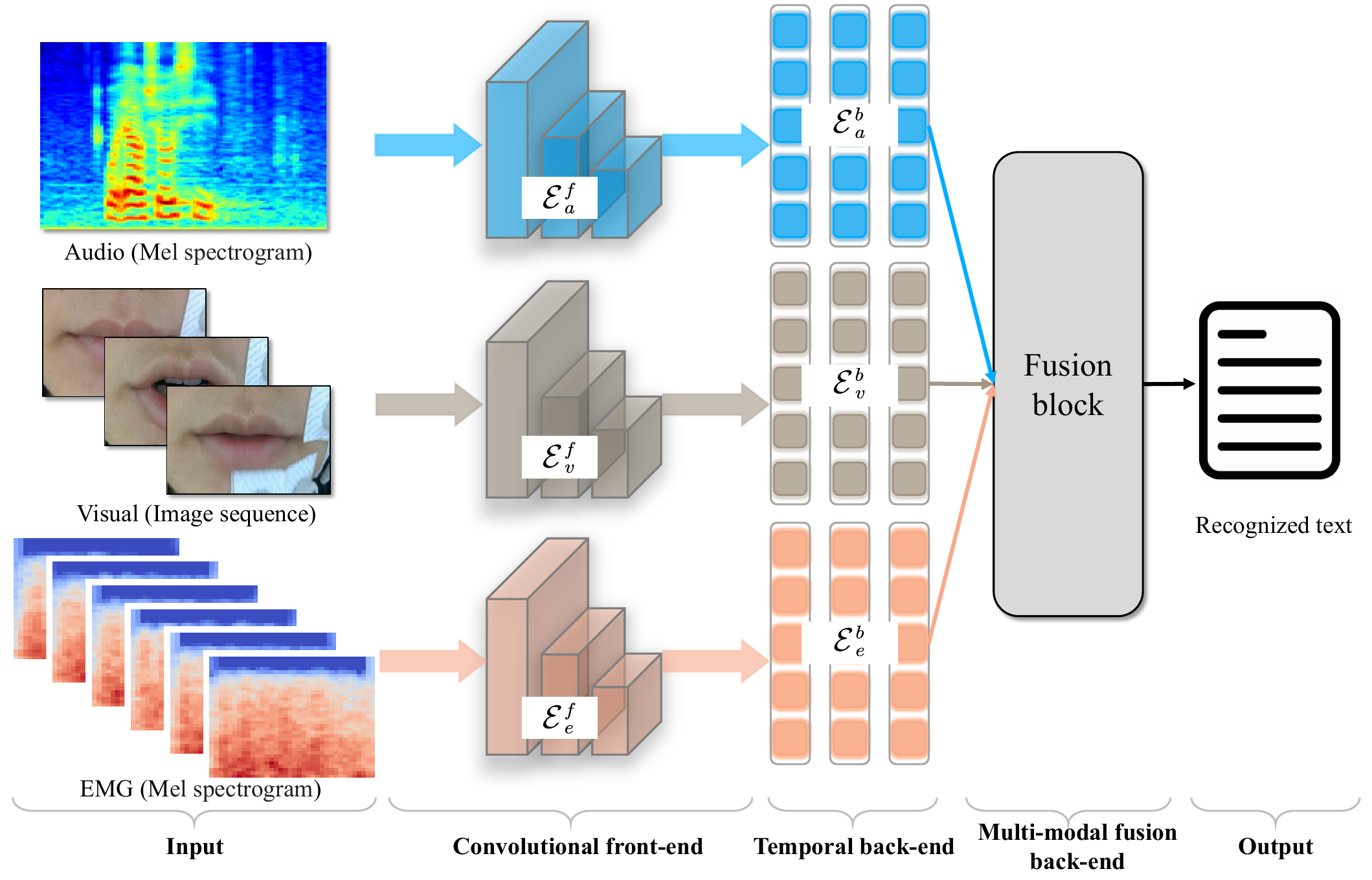}}
		\caption{Fusion network architecture. This fusion pipeline integrates three modalities of speech data: audio (Mel spectrogram), visual (image sequence of lip movements), and EMG (electromyography Mel spectrogram). For each modality, the input first passes through a convolutional front-end to extract spatial features, followed by a temporal back-end to capture temporal dependencies. The processed features from all three modalities are then fused in a multi-modal fusion back-end, which combines the complementary information to produce the final recognized text output.}
		\label{fig:fusion-pipeline}
	\end{figure*}
	
\textbf{Feature Processing.} Mel spectral features are extracted to capture acoustic characteristics from the perspective of human auditory perception. Building on the successful application of Mel frequency cepstral coefficient (MFCC) in automatic speech recognition systems, log Mel-frequency spectral coefficients (MFSC) are proposed, omitting the final discrete cosine transform (DCT) typically used in MFCC extraction. This approach preserves a higher-dimensional feature space. Accordingly, $60 \times  64$ MFSC features are extracted for each raw audio sample. Similarly, MFSC features with dimensions of $6 \times 36 \times 36$ are extracted as the basic feature set for the filtered EMG data.
	
	\section{Experiments}
	\label{sec:Experiments}
	In this section, we implement standard methods for single-modal speech recognition and employ conventional fusion paradigms to conduct multi-modal speech recognition experiments using the proposed dataset. An overview of the multi-modal speech recognition network utilized in this study is depicted in Fig. \ref{fig:fusion-pipeline}.
	
\subsection{Baseline Methods and Training Strategies}
\textbf{Single-Stream Speech Recognition Model.} For the audio and video feature streams, we adopt the front-end module developed in previous studies \cite{ma2021end}. The audio and lip-region image sequences are encoded using 2D and 3D convolutional layers, respectively, followed by the ResNet-18 model for feature extraction. A temporal back-end is integrated for temporal information processing. We employ a two-layer bidirectional gated recurrent unit (Bi-GRU) model and a Transformer encoder, chosen for their proven efficacy in previous research. The Bi-GRU layers have a hidden size of 512, while the Transformer encoder utilizes the default feed-forward network dimension of 2,048. For EMG-based silent speech recognition, we utilize a CNN architecture as a multi-channel spatial feature extractor. The front-end feature extraction for EMG data employs 2D convolutional layers with a kernel size of $3 \times 3$ and channel configurations of $[64, 64, 128, 128, 256]$. The same temporal back-end used in the audio and video streams is also applied to the EMG stream. The linear layer and SoftMax function are subsequently employed for sentence classification in the single-stream speech recognition model.

\begin{figure}[t]
		\centerline{\includegraphics[width=0.5\textwidth]{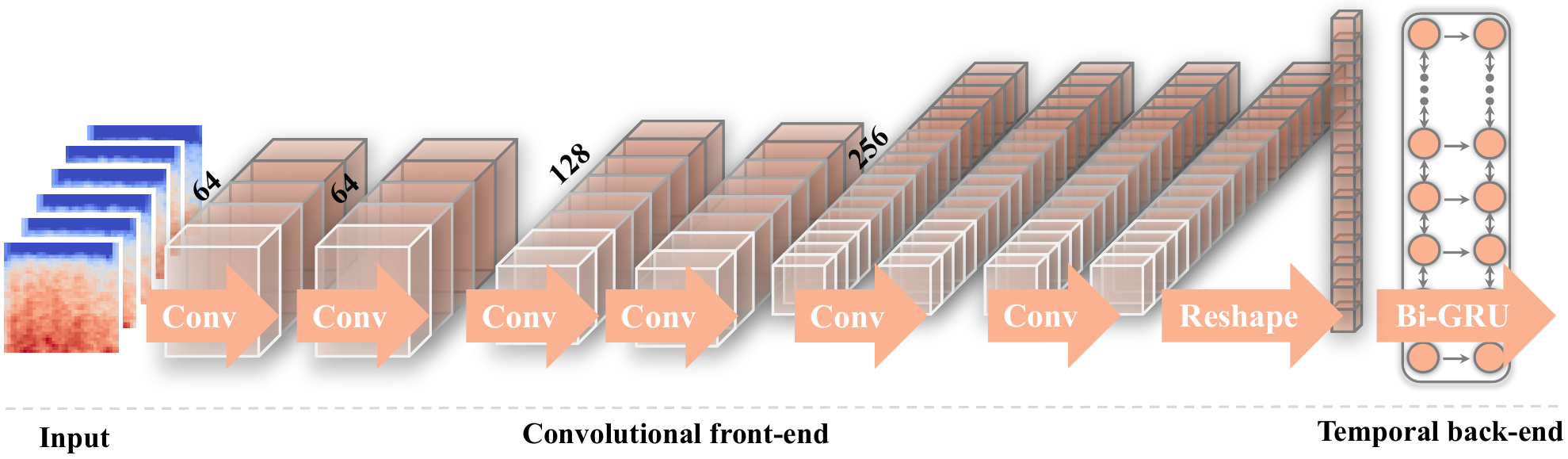}}
		\caption{Detailed architecture for the EMG branch.}
		\label{fig:emg-method}
	\end{figure}

\textbf{Multi-Modal Fusion Architecture.} We propose a multi-modal speech recognition network that integrates audio, visual, and EMG data based on the single-stream models outlined as Fig. \ref{fig:emg-method}. This network adopts feature fusion, one of the most widely used paradigms, which has demonstrated superior performance in audiovisual speech recognition tasks \cite{ma2021end}. The single-stream models act as spatio-temporal feature extractors for each modality, preserving unique speech information and passing it to the concatenation layer within the fusion block. Following feature concatenation, the fused multi-modal speech features are further processed by a temporal neural network for cross-modal temporal feature extraction, identical to the temporal back-end employed in each single stream. Notably, the Bi-GRU network is utilized as the temporal back-end in the feature fusion process due to its superior recognition performance observed in the single-stream experiments.
	
\textbf{Training Strategies.} To optimize model training and ensure convergence, we conducted pre-training on the three single-modal speech recognition models with the proposed dataset. The implementation was carried out using PyTorch, and the models were trained on the server equipped with an NVIDIA RTX 3090 GPU. The Adam \cite{kingma2014adam} optimizer was employed without weight decay. The initial learning rate was set to $3\times 10^{-4}$ for the Bi-GRU back-end, while a higher learning rate of $1\times 10^{-3}$ was applied to the Transformer encoder, with learning rate decay occurring after twenty to thirty epochs, depending on the modality. The batch size was set to 128 for both audio and EMG-based models, and 36 for the visual model. Training was guided by the cross-entropy loss function to evaluate model performance. Following pre-training, we froze the single-modal model parameters and loaded them into the multi-modal fusion back-end. This fusion model was then trained with a batch size of 64, while the number of training epochs varied depending on the specific modality. To support reproducibility and further research, all training configurations and source code are publicly available at: \url{https://github.com/MML-Group/code4AVE-Speech}.
	
	\subsection{Experimental Settings}
	\textbf{Cross-Subject Speech Recognition.} The entire AVE speech dataset was utilized for the cross-subject speech recognition experiments. Specifically, data from the first 70 subjects were used as the training set, data from the subsequent 10 subjects served as the validation set, and data from the remaining 20 subjects were allocated to the test set. Unlike prior research \cite{ma2021end}, which often involves training and test on data collected from the same speakers, this work targets speaker-independent speech recognition, closely aligning with real-world applications. As there was no overlap in subjects between the training and testing phases, the experiments evaluated the cross-subject recognition capabilities of the multi-modal fusion network. The recognition accuracy of each modality, as well as the fusion model, was reported for both the validation and test sets.
	
	\textbf{High-Noise Speech Recognition.} To further assess the robustness of the multi-modal fusion approach, we conducted experiments under high-noise conditions. 
	 Although the audio signals in the dataset were originally of high quality, Gaussian noise with varying intensities was added to the audio stream during the test phase to simulate acoustically challenging environments. Following the experimental setting described in \cite{stewart2013robust}, noise was introduced only during testing, without altering the training data.
	  The signal-to-noise ratio (SNR) was used as a metric to quantify the noise levels introduced into the test audio data, thereby enabling a thorough evaluation of the multi-modal fusion model's performance under adverse conditions.
	
    \subsection{Sentence-Level Recognition}
	\textbf{Single-Modal Speech Recognition Experiments.} To evaluate sentence-level speech recognition performance, we frame the task as a sentence-level classification problem and adopt accuracy as the evaluation metric. Accuracy is defined as:
    \begin{equation}
    \text{Accuracy} = \frac{1}{N}\sum_{i=1}^{N}  \mathbbm{1}(\mathbf{y}_i = \hat{\mathbf{y}}_i),
\end{equation}
where $N$ is the total number of samples, $\mathbf{y}_i$ is the true label for the $i$-th sample, $\hat{\mathbf{y}}_i$ is the predicted label for the $i$-th sample, and $\mathbbm{1}(\cdot)$ is a condition operation that gives one when the input is
true or zero when the input is false.
The recognition results of the single-modal speech recognition models are presented in Table \ref{tab1}, evaluated under varying SNRs during the test phase. For both the audio and EMG-based speech recognition models, the Bi-GRU back-end demonstrates superior recognition performance, whereas the lipreading model exhibits better results with an alternative configuration. Additionally, variations in the EMG signals, which are attributable to the physiological differences among subjects, impact the recognition performance. These variations lead to relatively lower performance in this cross-subject experiment. Given that EMG signals are minimally affected by environmental noise, they serve as a valuable complement to direct speech information when integrated with other modalities. As the SNR decreases, the recognition accuracy using the automatic speech recognition model declines significantly. This issue is exacerbated when the validation and test sets include previously unseen speakers, making it more challenging to maintain consistent recognition results. When the SNR falls below 0 dB, both EMG and visual modalities still provide sufficient speech information for recognition. This observation underpins the rationale for employing a multi-modal fusion approach to enhance speech recognition performance in this study.
	
	\begin{table}[t]
		\centering
		\renewcommand{\arraystretch}{1.3}
		\caption{Recognition accuracy of single-modal speech recognition models under varying noise conditions. Here, the term `back-end' refers to the temporal back-end architecture used in the single-modal branch, with no inclusion of multi-modal fusion components}
		\label{tab1}
		\resizebox{\linewidth}{!}{
			\begin{tabular}{lcccc}
				\toprule
				\textbf{Modality}    & \textbf{Back-end}           & \textbf{SNR of audio data} & \textbf{Validation set} & \textbf{Test set} \\ \midrule
				\multirow{12}{*}{Audio} & \multirow{6}{*}{Bi-GRU}        & Clean               & \textbf{99.47}          & \textbf{99.45}       \\
				&      & 10 dB              & 97.53          & 98.70       \\
				&       & 5 dB              & 91.65          & 95.40       \\
				&       & 0 dB              & 70.82          & 80.45       \\
				&      & -5 dB              & 36.41          & 45.47       \\
				&     & -10 dB             & 6.32          & 8.78       \\ \cline{2-5} 
				& \multirow{6}{*}{Transformer encoder} & Clean               & \textbf{99.30}          & \textbf{99.36}       \\
				&        & 10 dB              & 96.92          & 98.49       \\
				&        & 5 dB              & 86.98          & 93.98       \\
				&        & 0 dB              & 59.56          & 72.16       \\
				&       & -5 dB              & 28.61          & 35.33       \\
				&       & -10 dB             & 6.31          & 8.13       \\ \midrule
				\multirow{2}{*}{EMG}  & Bi-GRU                 & --               & \textbf{77.53}          & \textbf{75.53}       \\
				& Transformer encoder         & --              & 61.34          & 63.54       \\ \midrule
				\multirow{2}{*}{Visual} & Bi-GRU                 & --               & 92.45          & 95.09       \\
				& Transformer encoder         & --              & \textbf{98.71}          & \textbf{98.55}       \\ 
				\bottomrule
		\end{tabular}}
        \begin{tablenotes}
            {\footnotesize
            \item \textit{Note:} All accuracy values are reported as percentages, with the `\%' symbol omitted for clarity.}
        \end{tablenotes}
	\end{table}
	
	\textbf{Multi-Modal Speech Recognition Experiments.} Multi-modal fusion network experiments were conducted under various audio SNR conditions, as depicted in Table \ref{tab3}. The integration of audio, visual, and EMG modalities within the fusion model yields a marked improvement in recognition accuracy across different noise levels compared to the audio-only model. Notably, when the Transformer encoder is utilized for cross-modal temporal feature extraction, the recognition accuracy of the fusion model exhibits a slower degradation, maintaining an accuracy above 90\% even when the test audio SNR drops to -10 dB. 
    However, an intriguing counter-intuitive phenomenon is observed in conditions involving clean audio alone. In particular, the performance of the multi-modal fusion model slightly deteriorates compared to the model trained solely with clean audio, as highlighted in Tables \ref{tab1} and \ref{tab3}. We attribute this decrease to the additional complexity introduced by the multi-modal fusion network. Furthermore, our strict cross-subject experimental setup, where no speaker data overlap occurs between training and evaluation splits, amplifies the challenges in generalizing multi-modal models when only the clean speech modality is present during inference. Contrastingly, prior works \cite{prajwal2022sub,ma2023auto} often employ non-cross-subject splits, potentially masking such generalization issues.
    Nonetheless, the discrepancies in recognition performance for unseen speakers in the validation and test sets are significantly reduced compared to the single-modal models, highlighting the robust recognition capability of the fusion model for new speakers.
	
	\begin{table}[!t]
		\centering
		\renewcommand{\arraystretch}{1.2}
		\caption{Recognition accuracy of the multi-modal fusion network under varying noise levels. Here, the term `back-end' refers specifically to the back-end module of the multi-modal fusion block. For the single-modal branches utilized in the multi-modal fusion network, the Bi-GRU architecture is employed by default}
		\label{tab3}
		\resizebox{\linewidth}{!}{
			\begin{tabular}{lcccc}
				\toprule
				\textbf{Modality}    & \textbf{Back-end}          & \textbf{SNR of audio data} & \textbf{Validation set} & \textbf{Test set} \\ \midrule
				\multirow{12}{*}{Multi-modal} & \multirow{6}{*}{Bi-GRU}        & Clean              & \textbf{98.82}          & \textbf{99.18}       \\
				&       & 10 dB              & 98.46          & 99.01       \\
				&     & 5 dB              & 97.74          & 98.54       \\
				&       & 0 dB              & 95.93          & 97.17       \\
				&       & -5 dB              & 92.52          & 94.36       \\
				&      & -10 dB             & 86.41          & 88.57       \\ \cline{2-5} 
				& \multirow{6}{*}{Transformer encoder} & Clean              & \textbf{99.70}          & \textbf{99.87}       \\
				&        & 10 dB              & 99.58          & 99.81       \\
				&        & 5 dB              & 99.12          & 99.64       \\
				&        & 0 dB              & 98.15          & 99.12       \\
				&       & -5 dB              & 96.12          & 97.81       \\
				&      & -10 dB             & 93.00            & 95.51         \\ 
				\bottomrule
		\end{tabular}}
	\end{table}
	

    \subsection{Recognition Performance for Single Subject}
\begin{figure}[t]
		\centerline{\includegraphics[width=0.5\textwidth]{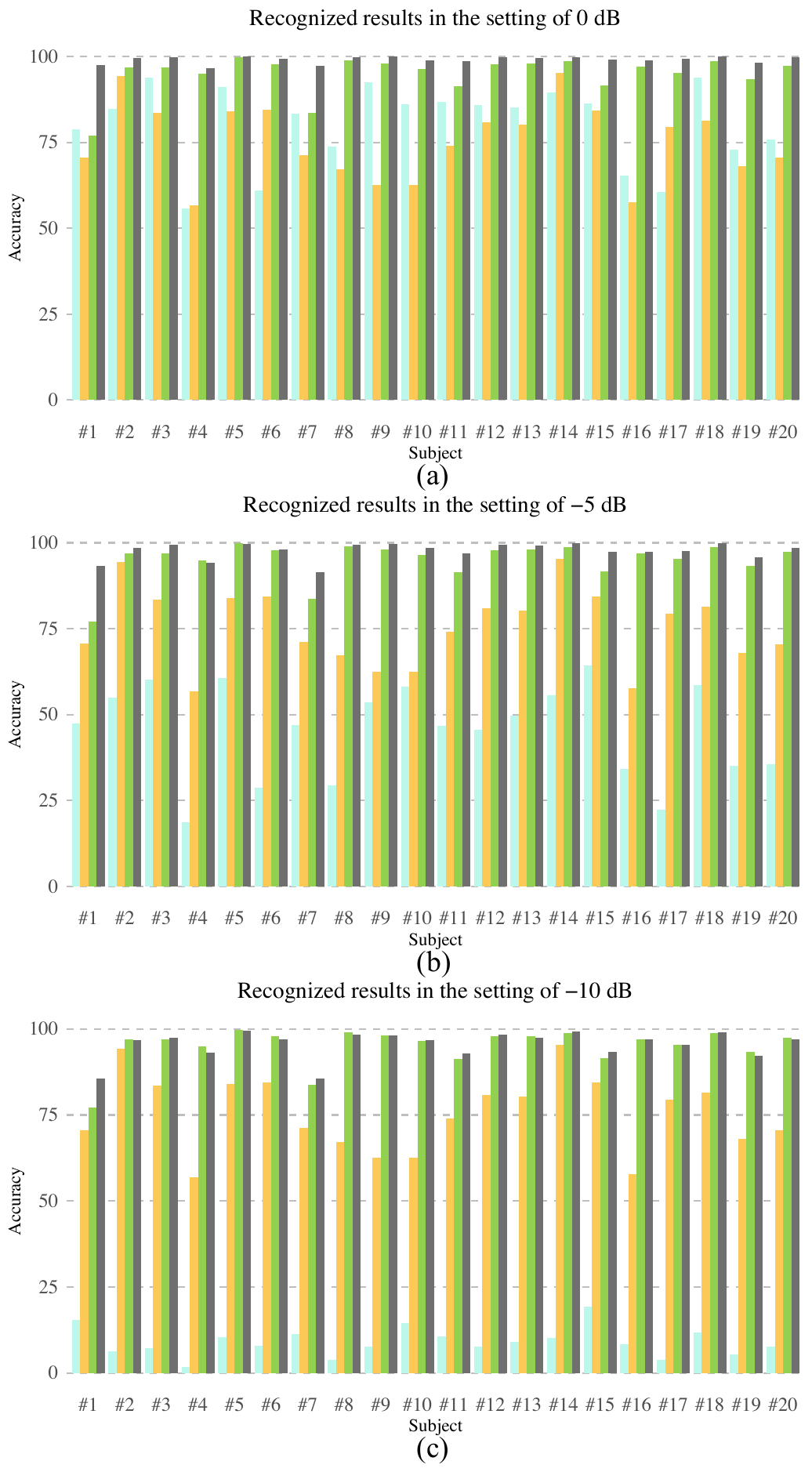}}
		\caption{Recognition results for unseen subjects using different modalities and the fusion network under varying levels of audio SNR at (a) 0 dB,  (b) -5 dB, and (c) -10 dB (here, ``\protect\begin{tcolorbox}[colback=color4audio, colframe=white, width=0.3cm, height=0.3cm, boxrule=0mm, halign=center, valign=center, nobeforeafter]\end{tcolorbox}'', ``\protect\begin{tcolorbox}[colback=color4emg, colframe=white, width=0.3cm, height=0.3cm, boxrule=0mm, halign=center, valign=center, nobeforeafter]\end{tcolorbox}'', ``\protect\begin{tcolorbox}[colback=color4vis, colframe=white, width=0.3cm, height=0.3cm, boxrule=0mm, halign=center, valign=center, nobeforeafter]	\end{tcolorbox}'', and ``\protect\begin{tcolorbox}[colback=color4fus, colframe=white, width=0.3cm, height=0.3cm, boxrule=0mm, halign=center, valign=center, nobeforeafter]	\end{tcolorbox}'' are the audio, EMG, visual, and the fused modalities, respectively. Zoom in for a better view).}
		\label{fig:fusion-three-modal}
	\end{figure}

    The sentence-level recognition results for each subject in the test set are depicted in Fig. \ref{fig:fusion-three-modal}, illustrating performance under varying levels of audio SNR at 0 dB, -5 dB, and -10 dB. Notably, there is significant physiological and appearance variation among subjects, which contributes to fluctuations in recognition accuracy. This variability is particularly pronounced when utilizing single-modal data, where the differences in recognition performance are more marked compared to the multi-modal fusion network. The experimental findings demonstrate that the integration of multi-modal speech data not only enhances recognition performance under conditions of severe audio noise but also mitigates the variability caused by subject differences. This effect is especially evident when the audio SNR decreases to -10 dB, as shown in Fig. \ref{fig:fusion-three-modal}(c). This outcome is critical for practical applications, where it is imperative that new users experience stable interaction without the need for model recalibration.
	The multi-level and rich-grained semantic information provided by the combination of audio, visual, and  EMG data effectively compensates for the limitations observed in single-modal speech recognition systems, particularly in cross-subject and high-noise scenarios. However, it is important to note that conventional fusion paradigms may exhibit inferior performance compared to single-modal streams when one of the modalities is severely compromised. For instance, when the SNR of the test audio data is reduced to -10 dB, the fusion network achieves recognition accuracies of 93\% and 95.51\% on the validation and test sets, respectively, using a Transformer back-end. These results are lower than those obtained by the visual speech recognition network, which achieves accuracies of 98.71\% and 98.55\% on the validation and test sets, respectively, using the same temporal back-end. 
	These observations suggest that there is a need for advanced fusion methods that can more effectively leverage the diverse modalities of speech information during the recognition process and resist noise interference in individual modalities.
	
\subsection{Prospective Study}

	The proposed multi-modal AVE speech dataset offers a comprehensive, high-quality, and synchronized speech dataset, collected from a diverse pool of speakers. This dataset serves as a critical resource for a broad spectrum of research and practical applications. Below, we outline several potential research directions, while acknowledging that the dataset may inspire numerous additional studies.
	
	\textbf{Multi-Modal Fusion Strategy.} Traditional fusion strategies, such as feature fusion \cite{ma2021end}, decision fusion \cite{yu2021fusing}, and their combinations, have been extensively studied in the field of multi-modal speech recognition. Additionally, various learning paradigms, including self-supervised learning \cite{shi2022learning}, have been employed to enhance the representational capacity of individual modalities and modality-agnostic linguistic features. The AVE multi-modal speech dataset, encompassing three distinct speech modalities, presents a unique opportunity to explore the representation of single-modal speech information and the correlations across modalities. The fusion process must address the differing feature dimensions and the sparsity of semantic information present in one-dimensional signals (e.g., audio and EMG) and high-dimensional image sequences (visual data). With the substantial volume of multi-modal data provided, researchers can develop both supervised and self-supervised learning approaches, as well as explore multitask learning and cross-modal knowledge distillation techniques, among others.

	\textbf{Speech Enhancement.} Research in multi-modal speech enhancement primarily focuses on improving speech quality and intelligibility \cite{zhang2020deepmmse,michelsanti2021overview}. Recent advancements in this area have largely utilized audio and visual speech information, leveraging methods such as disentanglement learning, generative modeling \cite{chou2024av2wav}, and unsupervised learning \cite{sadeghi2021mixture}. The multi-modal AVE speech dataset enables further investigation into the potential of using audio and EMG data for speech enhancement. Given that EMG signals directly reflect muscle movements during speech, they provide valuable insight into the relationship between articulatory motion and speech content. The similar waveform patterns between audio and EMG data offer a strong foundation for mutual information extraction, particularly in scenarios where audio signals are corrupted by noise or when speaker variability is present. Moreover, the integration of audio and EMG data can enhance lipreading performance under challenging conditions, such as high noise levels or poor lighting. By distilling knowledge from the other two modalities, single-modal speech enhancement techniques can achieve improved performance, enabling the use of a single modality in real-world applications.

\section{Limitations and Discussions}
\label{sec:limi_and_disc}
While the proposed dataset offers a valuable resource for advancing multi-modal speech recognition research, several limitations should be acknowledged. Primarily, the current participant group consists of younger adults with no known speech impairments. While practical and effective for establishing a high-quality baseline dataset, this recruitment approach may limit immediate generalizability to populations such as elder adults or individuals with speech disorders, who often experience different physiological and cognitive conditions affecting speech production and perception. Age-related changes—such as reduced respiratory capacity, decreased vocal fold elasticity, and muscular atrophy—can impact various aspects of speech, including fluency and articulation. Moreover, while our focus on visual lip movements and facial surface EMG signals is motivated by their relevance to individuals with vocal impairments, we acknowledge that the extent to which facial muscle activity patterns remain consistent across different populations remains an open area for future investigation. Despite these considerations, we believe the dataset serves as a robust foundation for the development of assistive and human-machine interaction technologies. It may also support comparative studies and serve as a reference point for models that can be adapted to more diverse user groups. In this context, transfer learning and domain adaptation approaches represent promising avenues for bridging demographic gaps. Future efforts could further strengthen this line of research by incorporating participants from a wider range of age groups and clinical profiles. Doing so would enhance the dataset’s ecological validity and ensure its utility across a broader spectrum of real-world scenarios and end-user needs.

	\section{Conclusion}
	\label{sec:cnc}
	In this paper, we introduced the AVE speech, a novel audio-visual-EMG-based multi-modal dataset specifically designed for Mandarin
    sentence-level speech recognition research. 
    The AVE speech dataset includes a Mandarin corpus of 100 classes of short sentences, collected from 100 subjects, providing a comprehensive resource for exploring multi-modal speech recognition.
	Using this dataset, we conducted experiments to compare the performance of multi-modal fusion models with single-modal speech recognition models under varying SNRs in the audio stream. The experimental results demonstrate that the fusion network significantly enhances recognition accuracy and robustness, particularly for unseen speakers. These findings suggest considerable potential for developing advanced fusion techniques that could further improve recognition performance in complex and challenging environments.
    The AVE Speech dataset offers a platform for future research into speech recognition and multi-modal learning and can serve as an essential tool for investigating the underlying mechanisms of human speech perception and interaction.

	\bibliographystyle{IEEEtran}
	\bibliography{references}

\end{document}